# A New Index based on Power Splitting Indices for Predicting Proper Time of Controlled Islanding

Hamzeh Davarikia
McNeese State University
Department of Electrical Engineering
Lake Charles, LA, USA
hdavarikia@gmail.com

Faycal Znidi
Department of Electrical Engineering
Texas A&M Uuniversity
Texarkana, TX, USA
fznidi@tamut.edu

Masoud Barati
Department of Electrical & Computer Engineering
University of Pittsburgh
Pittsburgh, PA, USA
masoud.barati@pitt.edu

Heena Rathore
Department of Computer Science
Texas State University
San Marcos, TX, USA
heena.rathore@ieee.org

*Abstract*— In the event of large disturbances, the practice of controlled islanding is used as a last resort to prevent cascading outages. The application of the strategy at the right time is crucial to maintaining system security. A controlled islanding strategy may be deployed efficiently at the right time by predicting the time of uncontrolled system splitting. The purpose of this study is to predict the appropriate islanding time to prevent catastrophic blackout and uncontrolled islanding based on existing relationships between coherent generator groups. A new instability index is derived from the proximity of inter-area oscillations to power splitting indices. Power splitting indices are derived using synchronization coefficients, which recognize the conditions in the system that warrant controlled islanding. The critical values of indices are calculated in offline mode using simulation data from IEEE 39-Buses, and their online performance is evaluated following a controlled islanding strategy. Through the introduction of these indices, system degradation can be effectively evaluated, and blackouts can be predicted early and prevented by controlled islanding at the right time.

## I. INTRODUCTION

Power systems are often subjected to large disturbances and are particularly critical at sites where the public safety and the environment are at risk. There are many different causes of large-scale power failures such as cascading failure, natural weather phenomena, short circuits, power surges, deterioration in substations, transmission lines, or other parts of the power distribution system. A power blackout is a short-term or a long-term loss in an area of the electric power and is the most critical form of the power outage that can arise, as well as the most difficult disruptions to recover from. Intentional Controlled Islanding (ICI) during the initial state of cascading outages development is frequently conducted as the final resort to preserve the power grid from severe blackouts in the event of the loss of synchronism. A strategy for controlled power system islanding is essential to the recent power defense system. It annihilates severe oscillations and disturbances by containing the impact of the disturbance to a small-island effect under different network configurations and system operating conditions in which the resilience of service would be slightly degraded [1]. The time at which each of the islands can be split from the system is critical for the success of the islanding condition, since early islanding or the late islanding may lead to an islanding indispensable dynamic behavior. Thus, it is essential to obtain an initial recognition that could designate if a disturbance will develop into a blackout or not, for predicting blackout occurrence. Whether, the islanding is triggered too late or far too early, a stable system can be forced to mistakenly be split into multiple islands, or an unstable system permitted to function and perhaps lead to uncontrolled separation or cascading blackout consequently.

The unstable inter-area oscillation can trigger out-of-step between two oscillating areas leading to system split. Concerning controlled islanding as a preventive defense versus unwanted system separation, two issues should be addressed: 1) "when to island" when to apply controlled islanding [1,2]. 2) "where to island" where are the proper areas for islanding [3–5]. For the techniques dealing with the problem "when to island" developing fasts algorithms based on the online measurement data using WAMS [6–8] and PMUs [9,10] are essential. In [11], a suitable time of controlled islanding using two oscillation zones with a generalized single machine in finite bus system (GSMIB) model was modeled. The equivalent power-angle curve is used for deriving critical power point and unstable equilibrium points-based which new instability indices (unstable equilibrium point IUEP) associated with inter-area oscillation are proposed. In [12], Using a prony method, timing in a controlled islanding strategy scheme that monitored the stability of the power system was presented. When a global instability in a power system is observed, the islanding signal was generated, and controlled islanding was done using the R-Rdot relays. In [13], a decision tree is used to determine the appropriate islanding time. However, the bus voltage angles are used for prediction and a list of critical events, which have the potential of a leading system to blackout is prepared in offline mode. In [14], a controlled islanding algorithm that uses spectral clustering over multilayered graphs where the multi-criteria objective function involves the correlation coefficients between bus frequency components and minimal active and reactive power-flow disruption to find a suitable islanding solution.

This paper proposes, a new index based on Power Splitting Indices (PSI) to predict the appropriate time of uncontrolled system splitting. At first, in offline mode coherent groups of generators based on the synchronization coefficients among generators are determined and the potential separation boundaries are optimized. Then, in the online monitoring, the separation boundaries are predicted by using the PSI. Finally, by calculating the risk threshold values of the PSI for each boundary between coherent groups the time to perform separation can be predicted. These separation indices are represented in three categories: a) the Incoherency of Generators in Different Coherent Groups (IGC) that depict to what degree the generators in different groups have the tendencies to swing contrary to the different groups; b) the Coherency of Generators in a Coherent Group (CCG), which depict in what way generators are coherently robust within the groups; and, c) the Detached Coefficient of Generators in Different Coherent Groups (DCGC), which unveils the system separation condition. By considering specific limited scenarios, the PSI's critical values that indicate the potential for possible uncontrolled islanding can be obtained. Simulation studies on the IEEE-39 test system are employed to demonstrate the effectiveness of the methodology.

## II. OVERVIEW OF THE PROPOSED METHOD

The proposed algorithm for assessing the appropriate time of controlled islanding is founded on the forecast of uncontrolled separation trend of a power grid in the event of a large disturbance based on the synchronization among the generators. Following disturbances, the coherency strength between generators in coherent groups gets altered. In particular, the coherency strength among generators in a coherent group increase, while the coherency strength decreases between different groups of the generators. The following section defines the procedure for identification of coherent groups of generators.

### A. Coherent group of generators

The coherent groups of generators determine based on the proposed approaches in [15], and [16]. Once, the number of coherent groups in any given system is determined, then inter-group relation with the symmetric matrix (KsGM) ($n \times n$) can be presented. The $KS$ matrix is partitioned to $u$ groups of coherent generators, which has an associated $u \times u$ matrix called KsGM. Fig. 1 shows the overall view of building the KsGM from the coherency coefficient between generators.

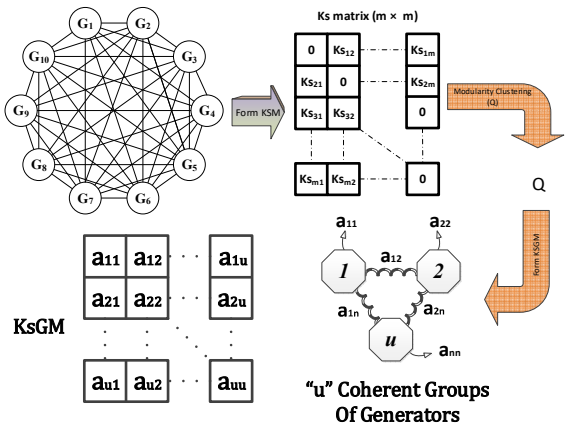

Fig. 1. Extraction of KsGM in an "m" machine system with "u" coherent group of generators

The square symmetric $u \times u$ matrix $KsGM = (KsGM_{ij} | i, j \in u)$ is the adjacency matrix associated with the complete graph of $u$ coherent groups of generators. The matrix has interesting properties that can be extracted by applying a simple algebraic process. The diagonal elements of matrix KsGM represent the strength of coherency in each group while the off-diagonal elements show how strong is the coherency among different groups [17-20]. Moreover, the Laplacian matrix and its eigenvalues present valuable information about the power system. Acquiring matrix KsGM in a real-time fashion paves the way for future analysis to observe the integrity of the power network. In this context, the proposed PSI based on the matrix KsGM will be introduced in the next section.

### B. Splitting indices for timing controlled islanding

To compute the proximity of inter-area oscillations to the point of system separation (out-of-step), three power splitting indices related to generator coherency are proposed for evaluating the tendency of oscillating areas for forced separation due to unstable inter-area oscillation. The Coherency of Generators in a Coherent group (CGC), is defined as the average of the diagonal entries of KsGM, which represents the strength of generator coherency within the groups.

$$CGC = \frac{1}{n}\sum_{i=1}^{n} a_{ii} \quad (1)$$

where $a_{ii}$ is the diagonal element of $KsGM$.

The Incoherency of Generators in Different Coherent Groups (IGC) is defined as the average of KsGM off-diagonal entries, which unveils to what extent the generators in different groups tend to swing against the other groups following a disturbance.

$$IGC = \frac{2}{n(n-1)}\sum_{i=1}^{n-1}\sum_{j=i+1}^{n} a_{ij} \quad (2)$$

The Detached Coefficient of Generators in different Coherent groups (DCGC) is defined as CGC divided by IGC, which represents the overall system separation status.

$$DCGC = \frac{n-1}{2} \frac{\sum_{i=1}^{n} a_{ii}}{\sum_{i=1}^{n-1}\sum_{j=i+1}^{n} a_{ij}} \quad (3)$$

For predicting the system probable for forced separation, it is essential to define a threshold for the value of power splitting indices before reaching a stable condition.

### C. Threshold of power splitting indices

In this section, the threshold of the PSI is found using a comprehensive list of input scenarios, including island and non-island conditions. The threshold values of the proposed method are calculated using different scenarios, which lead a group of generators to lose synchronism or to go out-of-step with the remaining generators. A 10-machine, the 39-bus test system is employed to demonstrate the PSI predictor scheme performance shown in Figure (2) below.

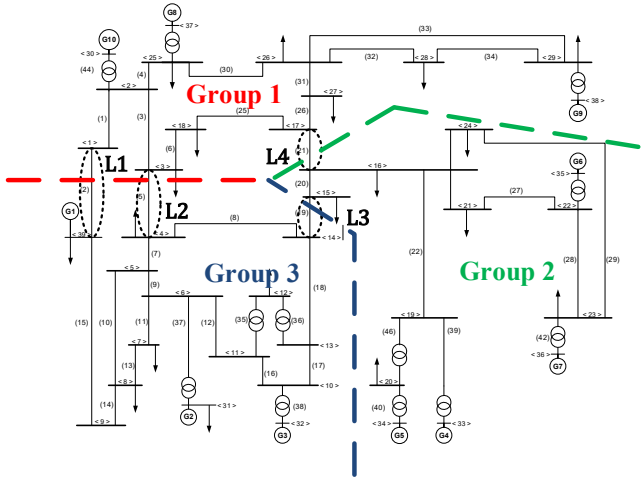

Fig. 2. IEEE 39-Buses system with groups of coherent generators boundary.

| System Status | Scenario | Event or Fault in Line (s) | Growing compared to the initial state | | |
|---|---|---|---|---|---|
| | | | % CGC | % IGC | % DCGC |
| Spontaneous Island Formation | 6 | 1&4(SwE) | 452 | 872 | 132 |
| | 7 | 2&3(SwE) | 251 | 660 | 78 |
| | 8 | 2&4(SwE) | 100 | 16285 | 162 |
| | 13 | 1&3(SC) | 93 | 251 | 44 |
| | 14 | 1 &4(SC) | 455 | 958 | 144 |
| | 15 | 2&3(SC) | 233 | 792 | 88 |
| | 16 | 2&4(SC) | 190 | 42132 | 173 |
| Machine Instability | 4 | 4 (SwE) | 110 | 1020 | 93 |
| | 5 | 1&3(SwE) | 354 | 256 | 68 |
| | 12 | 4 (SC) | 113 | 1095 | 95 |
| Stable with Low Swing | 1 | 1 (SwE) | 348 | 72 | 20 |
| | 2 | 2 (SwE) | 63 | 70 | 21 |
| | 9 | 1 (SC) | 351 | 75 | 21 |
| | 10 | 2 (SC) | 102 | 74 | 23 |
| | 3 | 3 (SwE) | 16 | 166 | 29 |
| | 11 | 3 (SC) | 38 | 177 | 32 |

TABLE I. THE CONSIDERED SCENARIOS FOR DETERMINATION OF THE THRESHOLD VALUES OF THE PSI

Previous studies show that disturbances can mostly affect the coherency with groups of coherent generators separated from other groups of coherent generators linked through weak interconnections. Such disturbances increase the generator's synchronism strength in a group and decrease relation in synchronism between different groups and lead to the loss of synchronism between the whole group. Accordingly, the threshold values of the PSI can be determined by creating disturbances between-group lines. Fig. 2 shows the lines 1-39 (L1) and 3-4 (L2) link coherent groups 1 and 3; the lines 14-15 (L3) link coherent groups 2 and 3 as well as lines 16-17 (L4) link coherent groups 1 and 2. Table 1shows all the considered 16 scenarios for the determination of the threshold values of the PSI on the connecting lines. The disturbances can be categorized into two groups, i.e. the switch-event (SwE) and the short-circuit Event (SC). Considering a switch event in one or two lines and a three-phase short-circuit fault at the location 50% along the length of the line happens in t=3 sec with clearing time of 100 msec. The system was then simulated for 5sec and at the end of each simulation, angular stability in the system was examined. The simulation is characterized as "stable with low swing" if all the machines stay in synchronism and their rotor angle has stable swings i.e., 6 scenarios. The simulation is discarded and labeled as "machine instability", i.e., 3 scenarios, if the isolated groups of machines exhibited loss of synchronism i.e., plant mode instability. The simulation characterized as "spontaneous island formation" which is fit for islanding, i.e., 7 scenarios, if the system spontaneously detached into obviously distinguishable groups of generators that have formerly been identified as coherent. Since the detection of uncontrolled islanding as well as the generator instability is the aim, the minimum value of the indices in the unstable group is considered. To this end, the threshold values for CGC, IGC, and DCGC are 93%, 251% and 44% are higher than normal conditions, respectively.

III. SIMULATION

To evaluate the effectiveness of the proposed indices methodology, the IEEE 39 bus test system was subjected to a wide assortment of disturbances not recognized by the indices for calculating PSI thresholds. Table 2 below shows the chosen cases.

TABLE II. THE CHOSEN CASES FOR THE ILLUSTRATION

| Scenario A | Scenario B |
|---|---|
| Switch event in line 3-4 and 13-14 | Three-phase short circuit in line 15-16 |

A. Scenario A

In this context, the performance of the proposed algorithm for the scenario of a 39-Bus test system at base case loading was subject to tripping in lines 3-4 and 4-14 at 3 sec and 10 sec respectively. As a result, lines 16-17 tripped at 15sec by its distance relay. These events caused a global instability, and coherent groups 1 and 2 start to separate from the system at t=25 sec and t=35 sec respectively as shown in Figure (3).

By applying the proposed strategy, the variation of the islanding indices CGC, IGC, DCGC, and the islanding signal throughout the blackout trend are illustrated in Figure (4), (a), (b), (c) and (d) respectively. As shown, CGC, IGC, and DCGC are increasing until reaching their critical values at 10 sec, and proposed algorithm signals were generated at 10.04 sec.

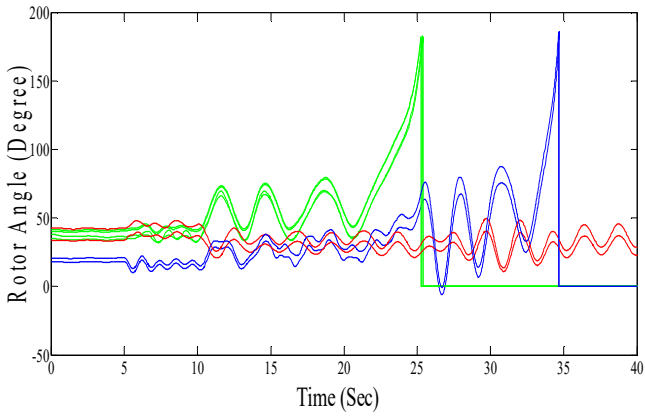

Fig. 3. Magnetization Rotor angle in test system without islanding

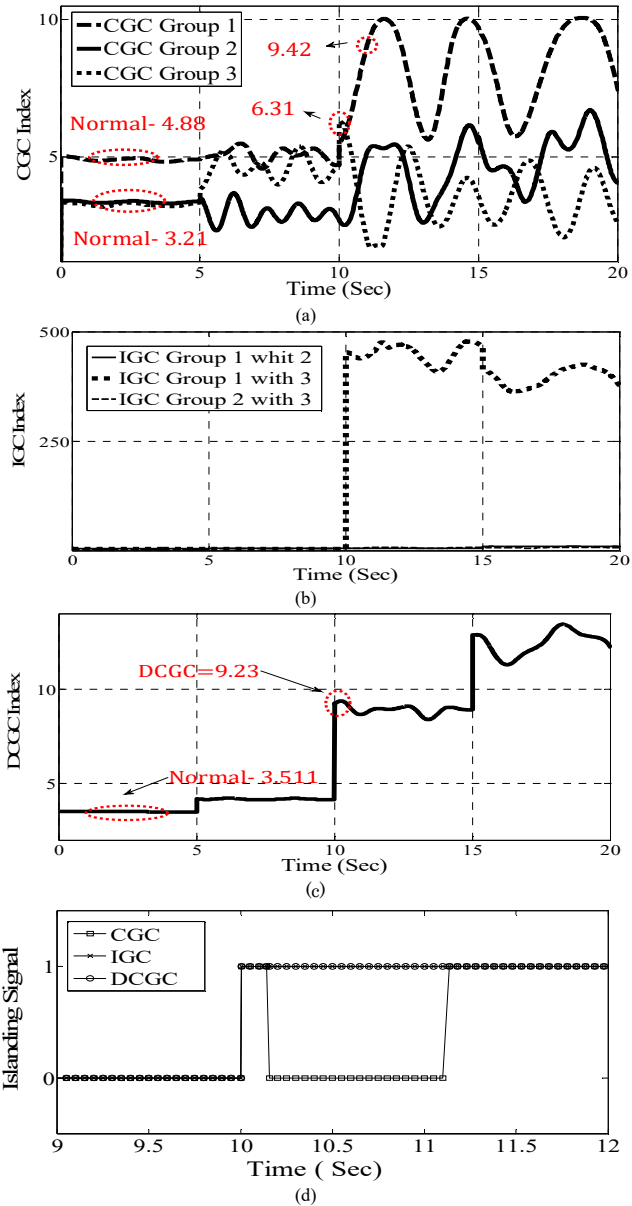

Fig. 4. (a-c) Variation of PSI in scenario A., (d) Variation of proposed indices signal in scenario A

The R-Rdot out-of-step relays were armed at 10.04 sec. It was seen that there is a clear change in lines resistance at t=10-12.5sec. Large unstable oscillations of the lines cause the apparent impedance property to fall inside the zone of the setting of the relay. Then, the R-Rdot relays elements operate during this power swing and detect the instability and cause interruption of the lines 14-15, 16-17 and 1-39 at 10.85sec, 11.31sec and 12.23 sec, respectively. Therefore, the frequency instability and voltage collapse can be prevented by the proposed indices by choosing the appropriate time as shown in Figure (5).

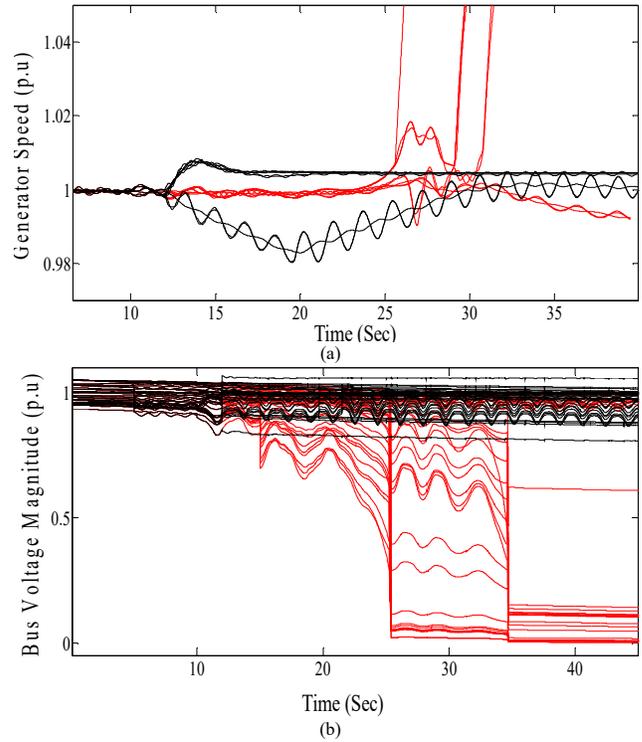

Fig. 5. (a) Generator speed with and without islanding. (b) Bus voltage magnitude with and without islanding in scenario A.

A comparison of the load losses without controlled islanding and with controlled islanding using proposed indices is shown in Figure (6). As can be seen, adopting the proposed algorithm would result in significant savings in load losses.

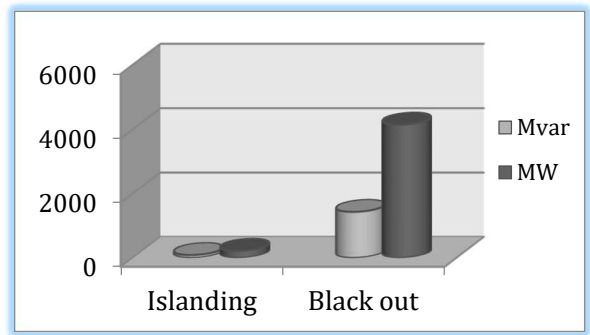

Fig. 6. Comparison of lost loads after blackout and islanding using the proposed method in scenario A

## B. Scenario B

In this case, a three-phase short circuit on line 15-16 with a fault period of 100 msec was simulated at 3 sec as a preliminary event. Cascading failures started by tripping the faulty lines 15-16 at 5 sec and 12 consequent lines tripped due to activation of distance relays. Then, two generators are removed at 5.6 sec.

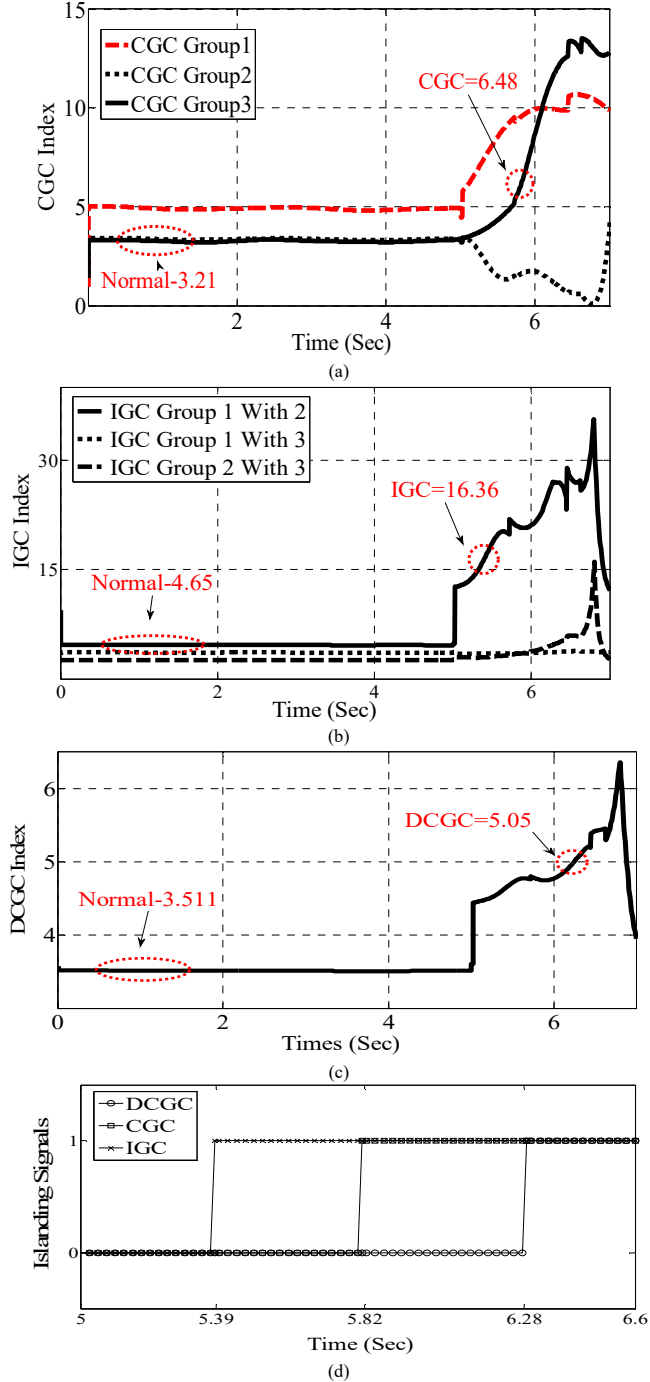

Fig. 7. (a-c) Variation of integrity indices in scenario B., (d) Variation of proposed indices signal in scenario B

The variation of islanding indices CGC, IGC, DCGC, and proposed indices signal during the blackout trend illustrated in Figure (7), (a), (b), (c) and (d) respectively. As shown, the CGC, IGC, and DCGC are increasing until reaching their critical values at 5.82, 5.03 and 6.29 sec respectively, and proposed indices signals were generated at 6.29 sec. A largely unstable oscillation of the lines causes the apparent impedance characteristic to fall inside the zone of the setting of the relay. Consequently, the R-Rdot relays elements operate during this power swing and detect the instability and cause interruption of the lines between 6.3-6.39 sec. table 3 shows the cascading outage in this scenario.

TABLE III. CASCADING OUTAGE WITHOUT ISLANDING IN SCENARIO B

| Event No | Place | Time(s) | Event No | Place | Time(s) |
|---|---|---|---|---|---|
| 1 | Line 15-16 | 5.00 | 9 | Line 08-09 | 8.20 |
| 2 | Line 02-03 | 5.72 | 10 | Line 26-27 | 8.74 |
| 3 | Line 17-18 | 6.44 | 11 | Line 09-39 | 8.92 |
| 4 | Line 04-05 | 6.44 | 12 | Gen 10 | 9.64 |
| 5 | Line 01-02 | 6.62 | 13 | Gen 09 | 9.95 |
| 6 | Line 03-04 | 7.16 | 14 | Line 26-28 | 10.68 |
| 7 | Line 13-14 | 7.16 | 15 | Line 26-25 | 10.68 |
| 8 | Line 04-14 | 7.17 | 16 | Load 39 | 19.99 |

As observed in Figure (8), the frequency instability and the voltage collapse could be prevented after the second event by the proposed indices by choosing the appropriate separation time. It is obvious that the proposed algorithm considerably reduces load shedding.

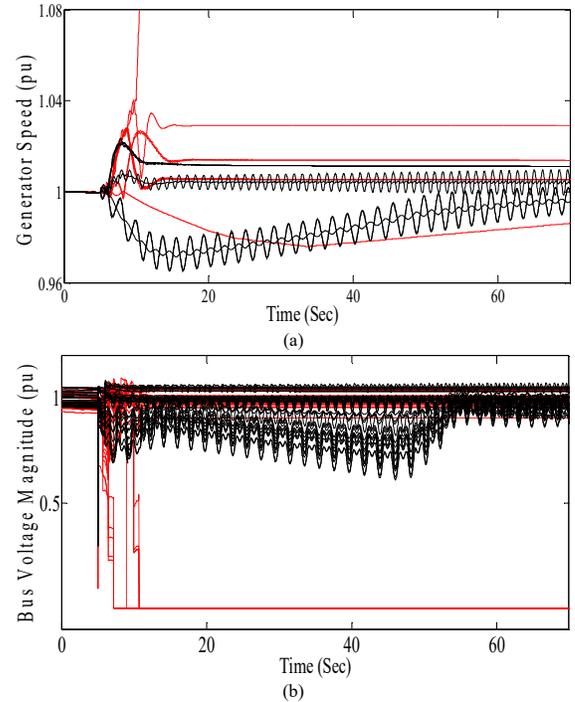

Fig. 8. (a) generator speed with and without islanding. 9b) bus voltage magnitude with and without islanding in scenario B

A comparison of the load losses without controlled islanding and with controlled islanding using proposed indices are shown in Figure (9).

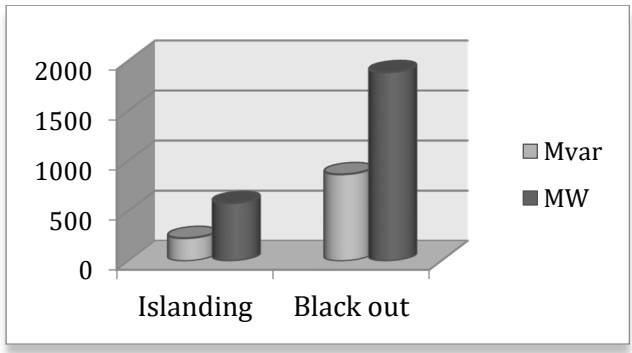

Fig. 9. Comparison of lost loads after blackout and islanding using proposed indices in scenario B

## IV. SIMULATION

The paper describes how synchronization coefficients-based indices can be used to detect cascading degradation of power system security at an early stage. By using classical transient stability modeling, synchronization coefficients among the system generators are introduced, and system splitting indices are used to evaluate system synchronism. Based on a comparison of synchronism between generators within each group, these indexes can predict uncontrolled islanding. Compared to other methods, the proposed method can be applied faster since there are fewer scenarios needed to determine the thresholds of indices. Further, the online value of indices can be calculated quickly using real-time grid data, and the decision to control islanding can be made instantly by observing the exceeding indices from the critical values. According to the results, the proposed method can effectively prevent cascading failures and blackouts.